\documentstyle[preprint,aps]{revtex} 

\begin{document}

\title{Electrical resistivity of a thin metallic film\/}

\author{Horacio E. Camblong$^{a}$
and Peter M. Levy$^b$}

\address{$^a$ Department of Physics, University of San Francisco, San
Francisco, California 94117 \\
$^b$ Department of Physics, New York University,
New York, New York 10003}

\maketitle


\begin{abstract}
The electrical resistivity of a pure sample of a thin metallic film 
is found to depend on the boundary conditions.
This conclusion is supported by a free-electron model calculation
and confirmed by an {\em ab initio\/} relativistic 
Korringa-Kohn-Rostoker computation.
The low-temperature resistivity is found to be
zero for a free-standing film (reflecting boundary conditions) 
but nonzero when the film is sandwiched between two semi-infinite samples 
of the same material (outgoing boundary conditions). In the latter
case, this resistivity scales inversely with the number 
of monolayers and is due to the background diffusive scattering by a finite
lattice.

\end{abstract}

\pacs{PACS numbers:
72.10.-d, 73.50.Bk, 73.50.-h, 75.70.Pa}


\section{Introduction}
\label{sec:introduction}

It is well known that the low-temperature electrical resistivity of an infinite
pure metal is essentially zero because the Bloch waves associated
with the underlying periodic structure undergo no diffusive scattering
in the absence of impurities and imperfections.
Then, one could ask what the corresponding result would be for
a thin metallic film. This question, which has not been addressed in the 
literature, is the central problem tackled in this paper. Our conclusion
is that the electrical resistivity of a thin film depends upon how
the experimental situation is set up, namely, upon the boundary conditions;
in particular, with an appropriate choice of boundary conditions,
the resistivity is not zero.

The problem of the electrical resistivity of a pure sample of a thin
metallic film is of great current interest.
It can be viewed as a fundamental question for which an exploratory
analytic calculation may provide deeper insight as well as a limiting
test of complex {\em ab initio\/} methods.
In fact, this question can be posed in the context
of the first-principle fully relativistic layered version of the 
Korringa-Kohn-Rostoker (KKR) method,
which has been recently developed~\cite{bla:98},
to perform {\em ab initio\/} computations of magnetotransport in 
magnetic multilayers in the coherent potential approximation.
In effect, this method can be easily applied to the computation
of the electrical resistivity of a thin film
consisting of a finite number $M$ of monolayers of a metal, 
by properly specifying the conditions enforced at the boundaries
of the system. The first result was obtained for a thin film of copper 
sandwiched between two 
semi-infinite metallic blocks of the same material;
the resistivity was computed for a film consisting of between 
$M=6$ and $45$ monolayers and for current in the plane of the layers (CIP) and
was found to decrease approximately as $116/M$ 
${\rm \mu\Omega \, cm}$, for $M>20$ 
ML---a result that clearly extrapolates to zero {\em only\/} as $M$ 
approaches infinity. This nonzero value is somewhat unusual and led us to
ask that the computation be done under different boundary 
conditions. For a free-standing slab, namely, when 
the copper film is surrounded by  vacuum on both sides, 
it was found that the resistivity is essentially zero~\cite{bla:99}.
In conclusion, when one performs these two computations with unequal
boundary conditions, one gets unequal resistivities.
In other words, the resistivity depends on the boundary 
conditions.

The results described above raise a number of questions
that we address in this paper.
Our goal is to explore electrical conduction in a pure sample of a thin 
metallic film and resolve the following fundamental issues.
(i) Is the resistivity indeed dependent on the boundary conditions?
(ii) How can one understand the finite resistivity of a thin,
yet otherwise perfect, film?
(iii)  How can one measure the finite resistivity of a thin film?

As we will see in Sec.~\ref{sec:bc}, a proper understanding 
of boundary conditions is a prerequisite for a thorough discussion of 
these questions.
This analysis will be followed in Sec.~\ref{sec:conductivity}
by the application of the Kubo formula to the case with outgoing
boundary conditions, in
Sec.~\ref{sec:periodicity} by the mathematical characterization of
the finite periodicity of the lattice, 
in Sec.~\ref{sec:resistivity} by the 
implications of the film finiteness 
on transport properties, 
and in Sec.~\ref{sec:conclusions} by concluding remarks.

\section{Resistivity and Boundary Conditions}
\label{sec:bc}

Electrical conduction in a thin metallic film can
be modeled by representing the film in terms of a finite number $M$ of
monolayers or perfect atomic planes arranged periodically in the 
direction perpendicular to the boundaries of the film.
Moreover, a pure sample is  characterized by the absence of
impurities or imperfections; in addition, at low temperatures,
other resistivity sources are rendered ineffective.

At first sight, one might just apply the standard folklore, namely,
that quantum-mechanical Bloch waves in a 
periodic structure propagate without electrical resistance.
In effect, electrical resistance  arises 
from the loss of linear momentum
information due to diffusive scattering, i.e., scattering 
that randomizes the electron's momentum, so that
the outgoing electron has no knowledge of
the direction of its incoming momentum~\cite{don:74}.
In fact, in a pure metallic sample, electrons undergo Bragg scattering, 
which being highly directional,
is ineffective as a momentum-randomizing mechanism
(as we will discuss in greater detail in Sec.~\ref{sec:resistivity}).
This would seem to imply that the resistivity of a perfect film
should be identically zero; however, this line of reasoning is 
simplistic: the standard folklore applies only to an {\em infinite\/} 
periodic sample, for which the electrical resistivity is indeed zero, 
under the conditions described above. Instead, when one analyzes a finite 
film, the question ``What is the resistivity of a thin metallic film?''
is immediately replaced by ``Is the film really periodic?''
Then, from a mathematical viewpoint, the film fails to be strictly 
periodic because of the boundaries; by abuse of language, one
could describe the film as exhibiting ``finite periodicity.''
It turns out that the finiteness of the film 
implies the existence of background diffusive scattering in addition to 
ordinary Bragg scattering (see Sec.~\ref{sec:resistivity}), 
and it is this diffusive scattering that becomes the source of a finite-size 
resistivity {\em if\/} the boundary conditions are appropriately selected.

The simplest boundary condition is provided by the ideal free-standing 
slab when a film with perfect boundaries is inserted in a vacuum.
Then, the electrons undergo specular reflections at 
the boundaries (with infinitely high potentials representing the 
onset of a vacuum) and effectively ``probe'' a truly periodic potential.
This amounts to repeating the film periodically; 
periodicity is restored by the boundary conditions and the 
standard folklore applies: the resistivity $\rho$ is indeed zero.
In fact, the absence of electrical resistivity can be
traced back again to the electrons keeping their memory of linear momentum.
In the old Fuchs-Sondheimer transport model~\cite{son:52}, 
this reflecting boundary condition, which amounts to the absence of diffuse 
scattering at the boundaries (100\% specular reflection), is parametrized 
by $p=1$, where $p$ is the coefficient characterizing the specularity of 
scattering off the surfaces of the film. The ensuing resistivity 
$\rho=0$ has been confirmed by the relativistic layered KKR 
method~\cite{bla:99}.

Does this mean that the resistivity is always zero?  Of course
not; the same relativistic layered KKR method showed that
finite-size effects are not negligible when the film
is sandwiched between two semi-infinite samples of the same material.
The novelty lies here in the use of different boundary conditions.
In effect, if the boundary conditions dictate that the electrons 
cannot keep their memory of momentum, then the background scattering
of the finite lattice becomes the source of a nonzero electrical 
resistivity.

An extreme form of this loss of momentum information is
achieved when the momentum of electrons entering the
finite sample is totally uncorrelated with that of electrons leaving 
the sample. One can conceptualize this extreme loss of momentum
information in three equivalent ways: in terms of reservoirs,
in terms of boundary scattering, and in terms of boundary conditions
for the electron propagators (Green's functions).
First, momentum loss is effectively implemented by having reservoirs that 
absorb the electrons upon leaving the sample---when the electrons probe 
a reservoir, they get out of synch with respect to their ``proper'' 
behavior in the sample.
Second, from the scattering viewpoint, this information loss
can be modeled by perfectly diffuse scattering at the boundaries of the 
film; this condition is precisely equivalent to the choice $p=0$ in the 
Fuchs-Sondheimer model~\cite{son:52} (0\% specular reflection, a condition
that totally erases momentum memory). The third viewpoint is needed
when applying the Kubo formula in terms of electron propagators;
in our original theory 
of transport in metallic superlattices~\cite{cam:95}, we described the 
required condition associated with this momentum loss as 
outgoing boundary conditions.

Transport theory 
with outgoing boundary conditions is based on the following ideas.
Transport is described via retarded Green's functions 
that represent the propagation of electrons in the environment 
provided by the sample, with the condition that the electrons leave the film 
irreversibly at the boundaries. This implies the use of Green's functions 
corresponding to an infinite medium (bulk)---at this level the 
calculation does 
not acknowledge the finiteness of the film whose conductivity is calculated. 
Then, as the propagators themselves do not satisfy boundary conditions 
that keep track of the momentum of electrons scattered within the film, 
the calculation yields a nonzero resistivity 
(see Sec.~\ref{sec:resistivity}). In other 
words, all finite conductors have self-energy terms in their
propagators that describe their contact with leads or reservoirs. The
resistance, which is proportional to the imaginary part of the self-energy,
reflects the fact that an electron in a finite conductor will eventually
leak out into the leads attached to it~\cite{dat:95}.

The discussion above has dealt successfully with the first two questions posed
in Sec.~\ref{sec:introduction}: the resistivity is indeed dependent 
on the boundary conditions
and we have understood conceptually how the finite resistivity of a 
perfect thin film arises under outgoing boundary conditions. 
However, the issue of how to measure the finite resistivity of a thin film 
has not yet been clarified. 
From the experimental viewpoint, the finite electrical 
resistivity of a perfect thin film still remains puzzling,
even when the concept of finite periodicity is introduced.
In effect, if we isolate the film from a bulk sample of copper and 
maintain it in contact with the ``remainder'' of the system (two perfectly 
conducting semi-infinite copper blocks), then, the  current is
shunted by these contacts. 
In other words, it would seem that it is not
possible to measure the resistivity of the film in this straightforward
way. In fact, one concludes that, for a measurement of electrical 
resistivity, the boundary contacts should not be of the same material.

So how does one observe the calculated resistance? 
As mentioned in the
previous paragraph, the shunting of the current by the contacts prohibits one
from measuring the resistance of a portion of a metal. 
However, if the current probe is sufficiently narrow 
to contact only the finite layer, 
and appropriate boundary conditions are enforced, then 
the current will be limited {\em only\/} to
the finite layer. 
Then, one can either measure the resistance of a free-standing film
or have it supported on an insulating substrate.
If the boundaries are ideal so that they have no appreciable roughness,
i.e., for $p=1$,
the boundaries simulate  reflecting or free-standing
boundary conditions and lead to zero resistivity.
Alternatively, if the boundaries are sufficiently roughened,
they simulate current flow {\em only\/} in the finite layer subject to
the $p = 0$ boundary condition---this amounts
to outgoing boundary conditions; in fact, the roughened interface could
even separate the finite layer from a semi-infinite
sample of the same material on either side, a situation that is modeled 
by the corresponding relativistic KKR computation~\cite{bla:99}.
 It is in this way that we understand the paradox
of using a film with a rough surface  to measure the resistance calculated
for a perfectly flat film; it is the boundary condition for transport on the
surface of the film that happens to be the same in both cases.

A parenthetical remark is in order.
The statement that $p=0$  corresponds to  
outgoing boundary conditions should not signify that the
resistivity coming from the ``bulk'' of the sample is directly related to the
scattering at the boundaries; the latter  are there merely to simulate the
boundary conditions that enable the ``bulk'' scattering to produce 
resistance. In other words, the surface should have  a roughness profile 
sufficient to guarantee the boundary condition $p= 0$; however,
further increasing the amplitude of the roughness, while increasing the 
resistivity due to the surface scattering, does not increase the resistivity 
coming from the bulk scattering. Therefore,
for  the case at hand, the  actual scattering does not come from
randomly situated impurities but from the potential of the 
positively charged background ions that form a finite, but otherwise 
perfect, lattice. A perfect periodic lattice has no
resistivity, and it is easy to overlook the fact that electrons are
scattered by it, i.e., they undergo Bragg scattering. As we will show in 
Sec.~\ref{sec:resistivity}, a finite but otherwise perfect lattice scatters 
electrons for all momenta, with two main contributions: constructive 
interference or Bragg scattering as well as background diffusive 
scattering. 
It is the latter that leads to electrical resistivity and vanishes in the 
infinite-thickness limit; unlike the case of impurity scattering, 
its resistivity is inversely proportional to the thickness of the film.

In the remainder of this paper, we will show
the details of the calculation of the electrical resistivity
of a thin film using the Kubo formula within the free-electron model.
Specifically, we will show that outgoing boundary conditions do 
indeed imply the existence of a finite electrical resistivity. 
Remarkably, the free-electron result agrees in form and reasonably
well in magnitude with the one calculated {\em ab initio\/} by Blaas
{\em et al.}~\cite{bla:98,bla:99} for a perfect slab of copper embedded in 
copper.

\section{Kubo formalism for the conductivity of a thin metallic slab}
\label{sec:conductivity}

As discussed in Sec.~\ref{sec:bc}, a perfectly diffuse boundary 
amounts to outgoing boundary conditions, which are implemented with 
the corresponding infinite-medium retarded Green's functions.
As the scattering ultimately leading to resistivity is due to a 
potential $V({\bf r})$ with ``finite periodicity,'' 
we resolve the Hamiltonian in the form 
\begin{equation}
H=H_{0}+ V({\bf r})
\;  ,
\end{equation}
where the unperturbed Hamiltonian $H_{0}$
 corresponds to free electrons
characterized by the eigenfunctions $ \varphi_{\bf k}
= \Omega^{-1/2} \,
e^{i{\bf k} \cdot {\bf r}}$;
in all quantum-mechanical computations, we will use particle-in-a-box
normalization with finite volume $\Omega$.
However, given a thin film of cross-sectional
area $A$, thickness $L$, and volume $\Omega=AL$,
the in-plane dimensions 
(defining the area $A$) will be effectively  regarded as infinite,
whereas the finiteness of the perpendicular or ``longitudinal''
dimension $L$ will become the source
of finite-size effects; this longitudinal direction will
be chosen to correspond to the $z$ axis.
Accordingly, whenever appropriate, a generic  vector
 ${\bf V}$ will be resolved into its 
in-plane component ${\bf V}_{\parallel}$ and its longitudinal component 
${\bf V}_{\perp}= \hat{ {\bf z} } V_{z}$;
with this notation,
the energy of the free-particle or unperturbed state of 
momentum $\hbar {\bf k}$ is
\begin{equation}
\epsilon_{{\bf k}} = 
\epsilon_{{\bf k}_{\parallel} k_{z}} = 
\epsilon_{{\bf k}_{\parallel} } + \epsilon_{k_{z}}
\;  ,
\label{eq:energy_resolution}
\end{equation}
where
$\epsilon_{{\bf k}_{\parallel} } 
= \hbar^{2} {{k}_{\parallel} }^{2}/2 m  $
and
$ \epsilon_{k_{z}}
= \hbar^{2} k_{z}^{2} /2 m  $,
with $m$  being the effective electron mass.
Our calculation of electrical conductivity will be performed by 
applying perturbation theory within the framework of the Kubo
formula~\cite{kub:57,mah:90}.
The required potential matrix elements are
given by
\begin{equation}
V_{{\bf k} {\bf k}^{\prime} } 
 =
\frac{1}{\Omega}  \int d^{3}r \,
e^{-i ( {\bf k} - {\bf k^{\prime}} ) 
\cdot{\bf r} }
\, 
V({\bf r}) 
 =  \frac{1}{\Omega}
\,
\widetilde{V}
({\bf k}-{\bf k^{\prime} })
\;  ,
\end{equation}
where
$\widetilde{V}
({\bf k})$ is the Fourier transform of the potential.
 Then, to second order in perturbation theory,
 the diagonal momentum-space elements of the t matrix are
\begin{equation} 
t_{\bf k}(\epsilon) = 
V_{{\bf k}{\bf k}} + 
\sum_{\bf k^{\prime}}
 V_{{\bf k}{\bf k^{\prime}}} 
G^{0}_{{\bf k^{\prime}}} (\epsilon) 
 V_{{\bf k^{\prime}}{\bf k}} 
\; ,
\end{equation}
where $  G^{0} (\epsilon)  $ is the free-particle propagator.
In order to evaluate the Kubo formula for
the electrical conductivity, 
the negative imaginary part of the t matrix
$\Delta=
-{\rm Im} (t)$ is needed. 
Then, from the condition
$V^{*}_{{\bf k}{\bf k^{\prime}}}=
V_{\bf k^{\prime}
{\bf k}}
$, it follows that
\begin{eqnarray}
\Delta({\bf k}_{\parallel}, k_{z};\epsilon) 
=  \Delta ({\bf k}; \epsilon)
& = & 
-{\rm Im} \left[
t_{{\bf k}}(\epsilon) \right]
\nonumber \\
&  = &
\pi
\sum_{ {\bf k}_{\parallel}^{\prime}, k_{z}^{\prime} } 
{\cal J}( {\bf k} - {\bf k^{\prime}})
\delta(\epsilon -\epsilon_{ {\bf k}_{\parallel}^{\prime} }
-\epsilon_{k_{z}^{\prime}})
\;  ,
\label{eq:Delta}
\end{eqnarray}
in which the effect of the potential on the conductivity is now
summarized by the function
\begin{equation}
{\cal J}({\bf q}) 
= 
\frac{\mid
\widetilde{V}({\bf q})
\mid^{2} \!}{ \Omega^{2}  }
\;  ,
\label{eq:J}
\end{equation}
with ${\bf q}={\bf k}-{\bf k^{\prime}}$, and
 where the density of states
$\delta(\epsilon - \epsilon_{ {\bf k}_{\parallel}^{\prime} }
-\epsilon_{k_{z}^{\prime}})$ 
has been directly extracted from 
$-{\rm Im} [ G^{0}_{{\bf k}^{\prime}_{\parallel} 
k_{z}^{\prime}}(\epsilon)]/\pi$.

The Kubo formula~\cite{kub:57}
gives the zero-temperature in-plane (CIP) dc
conductivity in terms of the in-plane current-current correlation 
function, by means of the formal double-limit expression~\cite{mah:90} 
 \begin{equation} 
\sigma
= 
\lim_{\beta \rightarrow \infty}
\lim_{\omega \rightarrow 0}
\frac{1}{2\omega \Omega} 
\,
\left[
\int_{0}^{\beta} d \tau e^{i \omega \tau}
\left\langle T_{\tau} {\bf j}_{\parallel} (\tau)
\cdot {\bf j}_{\parallel} (0)
\right\rangle
\right]_{i \omega \rightarrow \omega + i 0^{+}}
\; ,
\end{equation}
which in the independent-electron approximation reduces 
straightforwardly to the familiar form
\begin{equation}
\sigma
= 
\frac{e^{2}}{\Omega }\sum_{{\bf k}_{\parallel} ,k_{z},s}
\frac{1}{2}
\left[
\frac{\partial
\epsilon_{{\bf k}_{\parallel} k_{z}}}{\partial ( \hbar { k}_{\parallel} )}
\right]^{2}
\frac{\hbar}{2
\Delta({\bf k}_{\parallel}, k_{z};\epsilon_{F}) 
}
\delta(    \epsilon_{F}-
\epsilon_{{\bf k}_{\parallel} }
-
\epsilon_{ k_{z}} )
\; ,
\label{eq:Kubo}
\end{equation}
where $s$ stands for the electron spin index.
Equation~(\ref{eq:Kubo})
can be conveniently rewritten by evaluating 
the group velocity as
 \begin{equation} 
\left[
\frac{\partial\epsilon_{{\bf k}_{\parallel} }}{\partial(\hbar
 {\bf k}_{\parallel} )}
\right]^{2}=\frac{2\epsilon_{{\bf k}_{\parallel} }}
 {m}
\;  ,
\end{equation}
and applying the in-plane continuum limit
 \begin{equation}
 \frac{1}{A}
\sum_{{\bf k}_{\parallel} ,s}
\longrightarrow 
2\int\frac{d^{2}{\bf k}_{\parallel} }{(2\pi)^{2}}
 = 
\int d\epsilon_{{\bf k}_{\parallel} }
g_{\parallel}(\epsilon_{{\bf k}_{\parallel} })
\;  ,
\label{eq:cont_in-plane}
 \end{equation}
with the only restriction that the integrand be
 a spin-independent function  
of the momentum variables
$ |{\bf k}_{\parallel}| $ and $k_{z}$
(but not of the corresponding angular in-plane variable).
In Eq.~(\ref{eq:cont_in-plane}),
 $g_{\parallel}(\epsilon_{{\bf k}_{\parallel} })$ is
the in-plane two-dimensional (2D) density of states per spin degree of 
freedom and per unit area, which 
is given by
\begin{equation}
 g_{\parallel}(\epsilon)
=\frac{1}{4\pi} \,
 \frac{2m}{\hbar^{2}}	
\;   ,
\label{eq:density_par}
\end{equation}
provided that the in-plane dimensions be effectively infinite.
Notice that this density of states 
$ g_{\parallel}(\epsilon)$ is actually a constant, $g_{\parallel}$,
which from now on will be factored out of the 
corresponding integrals. 
 Then, Eqs.~(\ref{eq:Kubo})--(\ref{eq:density_par}) 
imply that the conductivity is given by the expression 
 \begin{equation} 
\sigma
 = 
\frac{e^{2}}{ h} \, 
\frac{1}{L} \sum_{k_{z}}
\frac{  \epsilon_{F} - \epsilon_{k_{z}} }{\Delta 
\mbox{\boldmath\large  $\left(  \right.$ } \! \!
k (\epsilon_{F}- \epsilon_{k_{z}}), 
k_{z}; \epsilon_{F}
 \!  \mbox{\boldmath\large  $ \left.  \right)$ }
}
\;  ,
\label{eq:conductivity_sum}
\end{equation}
where $k (\epsilon)$  is the
positive root of  the equation  $\epsilon_{k}
= \epsilon$, i.e., 
$k(\epsilon)
= \sqrt{ 2m \epsilon}/ \hbar $; in particular,
\begin{equation}
 k ( \epsilon_{F} - \epsilon_{k_{z}} )
= \sqrt{ k_{F}^{2} - k_{z}^{2} }
\;  .
\end{equation}
In Eq.~(\ref{eq:conductivity_sum}), 
it is assumed that 
$\Delta({\bf k}_{\parallel}, k_{z}; \epsilon_{F}) $
is  a function of $|{\bf k}_{\parallel}|$ and $ k_{z}$ alone;
thus, to simplify the notation,
from now on this quantity will be represented as 
$\Delta(|{\bf k}_{\parallel}|, k_{z}; \epsilon_{F}) $.

Equation~(\ref{eq:conductivity_sum}) will be the starting
point for our conductivity calculation
in Sec.~\ref{sec:resistivity}, where
 we will straightforwardly apply its
counterpart in the longitudinal continuum limit,
 \begin{eqnarray}
 \frac{1}{L}
\sum_{k_{z}}
f \left(
 k_{z} 
\right)
&  \longrightarrow & \;
\int_{-\infty}^{\infty}
\frac{dk_{z}}{2\pi}
f \left( 
 k_{z} \right)
\nonumber \\
& = & \;
\frac{1}{2}
\int_{0}^{\infty} d
 \epsilon_{k_{z}} g_{\perp}(\epsilon_{k_{z}}) 
\,
\left[
f 
\mbox{\boldmath\large  $\left(  \right.$ } \! \!
   k (\epsilon_{k_{z}}) 
 \!  \mbox{\boldmath\large  $ \left.  \right)$ }
+
f 
\mbox{\boldmath\large  $\left(  \right.$ } \! \! \! \!
 - \! k (\epsilon_{k_{z}}) 
 \!  \mbox{\boldmath\large  $ \left.  \right)$ }
\! \right]
\;  ,
\label{eq:cont_longitudinal}
 \end{eqnarray}
where
\begin{equation}
g_{\perp}(\epsilon)=
 \frac{1}{2\pi} 
\left(
\frac{2m}{\hbar^{2}} \right)^{1/2} 
\, 
\epsilon^{-1/2}
\;  
\label{eq:density_perp}
\end{equation}
 is the longitudinal density of states per spin degree of freedom  
and per unit length;
notice that, explicitly,  $k (\epsilon_{k_{z}}) = |k_{z}|$. 
Under this approximation,
Eq.~(\ref{eq:conductivity_sum})
 turns into
 \begin{eqnarray} 
\sigma
 = 
\frac{e^{2}}{ 2 h } \, 
 \int^{\epsilon_{F}}_{0} 
d\epsilon_{k_{z} }
\,   &  &
\left( 
\epsilon_{F} - \epsilon_{  k_{z} } \right) 
\,  g_{\perp}(\epsilon_{k_{z} }) 
\nonumber
\\
&  & 
\mbox{}
 \times 
\left[
\frac{1}{ 
\Delta 
\mbox{\boldmath\large  $\left(  \right.$ } \! \!
 k (\epsilon_{F} - \epsilon_{k_{z}}), k (\epsilon_{k_{z}}) ;\epsilon_{F}
 \!  \mbox{\boldmath\large  $ \left.  \right)$ }
}
+
\frac{1}{
\Delta 
\mbox{\boldmath\large  $\left(  \right.$ } \! \!
 k (\epsilon_{F} - \epsilon_{k_{z}}), - k (\epsilon_{k_{z}}) ;\epsilon_{F}
 \!  \mbox{\boldmath\large  $ \left.  \right)$ }
}
\right]
\;  .
\label{eq:conductivity_integral}
\end{eqnarray}

\section{Finite Periodicity}
\label{sec:periodicity}

A thin film can be regarded as built out of
primitive cells assembled into an effectively infinite arrangement
in two directions (for which we will use symbols 1 and 2)
and a finite layering in a third direction (for which we will 
use the symbol 3).
As we will consider a generic ``finite Bravais lattice,'' 
the three directions need not be perpendicular to each other.
In other words, if the number 
of primitive cells stacked in direction $j$ is $N_{j}$, then 
$N_{1}, N_{2} \gg N_{3}$, and, in practice, we will regard 
$N_{1}$ and $ N_{2}$ as effectively infinite but 
$ N_{3}$ as a finite number; notice that
 $N = N_{1} N_{2} N_{3}$ is the total number of primitive cells in the 
film.
The finite periodicity in the third direction can be described by
counting the number $M=N_{3}+ 1 \approx N_{3}$ of stacked infinite 
lattice planes or monolayers; notice that, for the sake of 
simplicity, we will assume $M \gg 1$.
For instance, if the film has thickness $L$
 and consists of exactly $M$ monolayers from one 
boundary to the other, then  $L=N_{3} d \approx M d$,
with $d$ being the distance between consecutive monolayers.

The basic periodicity 
of the crystal structure within its boundaries
can be described by means of a finite generalization
of the concept of Bravais lattice.
As usual, given the primitive translation vectors
${\bf a}_{j}$, with $j=1,2,3$, it follows that the set of all
translation vectors
${\bf R}$ is of the form
${\bf R}= n_{1} {\bf a}_{1} + n_{2} {\bf a}_{2} + n_{3} {\bf a}_{3}$,
where the integers $n_{j}$ are limited to the values
 $ 0 \le n_{j} \le N_{j}-1$ for a finite lattice.
Then, for any local function of the position, such as the
lattice potential, the property
\begin{equation}
V({\bf r}+ {\bf R}) = V (  {\bf r})
\;  
\end{equation}
remains valid 
within the boundaries of the film.
In particular, for Fourier transforms, finite
periodicity guarantees the  identity
\begin{equation}
\int_{ T_{\bf R} {\cal C}  }  
 d^{3} r  \,
e^{-i {\bf q} \cdot {\bf r} } 
V ( {\bf  r})
 = 
e^{-i {\bf q} \cdot {\bf R}}
\int_{ {\cal C}}  
 d^{3} r  \,
e^{-i {\bf q} \cdot {\bf r}} 
V ( {\bf  r}) 
\;  ,
\end{equation}
where an arbitrary  primitive cell 
$ {\cal C}$ can be  related to
  any other primitive cell via a translation
$ T_{\bf R}  $ by a Bravais lattice vector ${\bf R}$.
This property leads to the characteristic Bragg peaks
with respect to electronic conduction, as shown below.
In effect,
the  Fourier transform of the potential becomes
\begin{eqnarray}
\widetilde{V}({\bf q})
& =  & 
\int_{\cal V}  d^{3} r  \,
e^{-i {\bf q} \cdot {\bf r} } 
V({\bf r})   
\nonumber \\
& = & \;
\sum_{n_{1}=0}^{N_{1}-1} 
\sum_{n_{2}=0}^{N_{2}-1} 
\sum_{n_{3}=0}^{N_{3}-1} 
e^{-i {\bf q} \cdot \left(
n_{1} 
{\bf a}_{1}  +
n_{2}  {\bf a}_{2}  +
n_{3} {\bf a}_{3}  \right)
} 
\int_{\cal C}  d^{3} r \,
e^{-i {\bf q} \cdot {\bf r} } 
V({\bf r}) 
\;  ,
\label{eq:periodFourier}
\end{eqnarray}
where ${\cal V}$ is the entire sample and  
${\cal C}$ is a 
reference primitive cell.
Then, summing the geometric progressions involved in 
Eq.~(\ref{eq:periodFourier}) and replacing in Eq.~(\ref{eq:J}),
one finds
\begin{equation}
{\cal J} ({\bf q})=
\left[ \prod_{j=1}^{3}
{\cal F}_{N_{j}}( {\bf q} \cdot {\bf a}_{j} ) 
\right]
\,
{\cal J}^{(0)} ({\bf q})
\;  ,
\label{eq:potentialsquared}
\end{equation}
where 
\begin{equation}
{\cal J}^{(0)} ({\bf q}) =
\left| 
\int_{\cal C}  \frac{d^{3} r }{v} 
e^{-i {\bf q} \cdot {\bf r}}
 V({\bf r}) 
 \right|^{2}
\;  
\label{eq:potentialsquared_0}
\end{equation}
is the corresponding quantity for just one primitive cell,
 $v={\bf a}_{1} \cdot ({\bf a}_{2} \times {\bf a}_{3})$
 is the volume of a primitive cell,
and 
\begin{equation} 
{\cal F}_{N_{j}}(\xi_{j}) =
\frac{1}{N_{j}^2} 
\left[
\frac{\sin ( N_{j} \xi_{j}/2 ) }{\sin (\xi_{j}/2)}
\right]^{2}
\label{eq:FN}
\end{equation}
stands for the interference factor associated with $N_{j}$ 
identical primitive cells aligned in the direction of
the primitive translation vector ${\bf a}_{j}$, with 
\begin{equation}
\xi_{j}= {\bf q} \cdot {\bf a}_{j}
\; .
\label{eq:xi_j}
\end{equation}
The interference factors are the origin
of the Bragg scattering peaks, which are represented by
$\delta$ functions and amount to the selection of the conditions 
$\xi_{j}={\bf q} \cdot {\bf a}_{j}   = 2 \pi n_{j}$ (with $n_{j}$ 
integer numbers),
as can be seen from the vanishing of the denominator
in Eq.~(\ref{eq:FN}).
These $\delta$ functions can be explicitly 
 displayed by means of the expansion in a series of simple fractions
\begin{equation}
\frac{1}{\sin^2 (\xi_{j}/2)} = \sum_{ n_{j} = -\infty}^{\infty}
\frac{1}{(\xi_{j}/2-n_{j} \pi)^2}
\;  
\label{eq:exp_simple_fractions}
\end{equation}
and from the familiar asymptotic result
\begin{equation}
\frac{1}{N_{j}}
\left[ \frac{ \sin(N_{j} \xi_{j}/2 )}{(\xi_{j}/2)} \right]^{2} 
\sim
\pi \delta(\xi_{j}/2)
\, 
\label{eq:Mlimit_aux}
\end{equation}
 for  ${\scriptsize N_{j} \rightarrow \infty}$;
then, the interference factors become 
\begin{eqnarray} 
{\cal F}_{N_{j}}(\xi_{j} ) & = & \;
 \frac{1}{ N_{j}^{2} }
\sum_{n_{j}=-\infty}^{\infty}
\left[
\frac{\sin (
N_{j} \xi_{j}/2)}{(\xi_{j}/2  - n_{j} \pi)} 
\right]^{2}
\nonumber \\
& 
\sim & \;
\frac{2\pi}{ N_{j}}
 \, \sum_{n_{j}=-\infty}^{\infty}
\delta(\xi_{j} - 2 \pi n_{j})
\;  ,
\label{eq:bragg_deltas}
\end{eqnarray}
where the second expression is to be understood as 
the asymptotic form for $N_{j}$ ``sufficiently large.''
The continuum limit implicit in Eq.~(\ref{eq:bragg_deltas}) can
be reversed by replacing the Dirac delta function by a Kronecker delta
and recalling Eq.~(\ref{eq:xi_j}), whence
\begin{equation} 
{\cal F}_{N_{j}}(\xi_{j})   
\sim
  \sum_{n_{j}=-\infty}^{\infty}
\delta_{ {\bf q} \cdot {\bf a}_{j}, 2 \pi n_{j}}
\;  .
\label{eq:bragg_deltas_discrete}
\end{equation}
In particular, the combination of the three
structure factors in Eq.~(\ref{eq:potentialsquared})
amounts to 
\begin{equation} 
\prod_{j=1}^{3}
{\cal F}_{N_{j}}(\xi_{j})   
\sim
  \sum_{n_{1}, n_{2}, n_{3} = - \infty}^{\infty}
\delta_{ {\bf q} \cdot {\bf a}_{1}, 2 \pi n_{1}}
\delta_{ {\bf q} \cdot {\bf a}_{2}, 2 \pi n_{2}}
\delta_{ {\bf q} \cdot {\bf a}_{3}, 2 \pi n_{3}}
\;  .
\label{eq:structure factor1}
\end{equation}
Equation~(\ref{eq:structure factor1}) can be reinterpreted by 
expanding ${\bf q}$ in terms of primitive reciprocal vectors,
i.e., ${\bf q}= \nu_{1} {\bf b}_{1}
+ \nu_{2} {\bf b}_{2} + \nu_{3} {\bf b}_{3}$,
with  ${\bf a}_{j} \cdot {\bf b}_{h} = 2 \pi \delta_{jh}$,
whence $\nu_{j} = {\bf q} \cdot {\bf a}_{j}/2 \pi$.
Then, Eq.~(\ref{eq:structure factor1}) 
states that $\nu_{j}=n_{j}$ is an integer, 
so that ${\bf q}$ is indeed a reciprocal-lattice vector
${\bf G}_{n_{1} n_{2} n_{3}} $, i.e.,
\begin{equation} 
\prod_{j=1}^{3}
{\cal F}_{N_{j}}(\xi_{j})   
\sim
  \sum_{n_{1}, n_{2}, n_{3} = - \infty}^{\infty}
\delta_{ {\bf q},  {\bf G}_{n_{1} n_{2} n_{3}}}
\;  .
\label{eq:structure factor2}
\end{equation}

However, in a finite lattice, it is legitimate to
apply the limit of Eq.~(\ref{eq:structure factor2})
{\em only\/} with respect to the directions defined by the 
primitive vectors ${\bf a}_{1}$  and  ${\bf a}_{2}$.
Then, the component of ${\bf q}$ on the plane spanned
by the primitive reciprocal vectors ${\bf b}_{1}$  and  ${\bf b}_{2}$
is
\begin{equation}
{\bf q}_{\rm in} =
{\bf q}  - \frac{ \left( {\bf q} \cdot {\bf a}_{3}  \right)}{2 \pi}
{\bf b}_{3} = {\bf G}_{n_{1}n_{2} 0} 
\;  ,
\end{equation}
which is a 2D reciprocal vector,
\begin{equation}
{\bf g}_{n_{1}n_{2}}
= {\bf G}_{n_{1}n_{2} 0} =
n_{1} {\bf b}_{1}
+ n_{2} {\bf b}_{2}
\;  ,
\end{equation}
so that
\begin{equation} 
\prod_{j=1}^{2}
{\cal F}_{N_{j}}(\xi_{j})   
\sim
  \sum_{n_{1}, n_{2} = - \infty}^{\infty}
\delta_{ {\bf q}_{\rm in},  {\bf g}_{n_{1} n_{2} }}
\;  .
\label{eq:structure factor3}
\end{equation}
Next we will assume that, for  ${\bf q} = {\bf k} - {\bf k}'$, 
with ${\bf k}$ and ${\bf k}'$ on the Fermi surface,
the only allowed 2D reciprocal-lattice vector
${\bf g}_{n_{1}n_{2}}$  is the zero vector.
To see that this assumption is reasonable,
let us consider, for example, a thin film of copper,
for which one can choose the set of primitive vectors
${\bf a}_{1}= (a/2) \, (\hat{ \bf x }  -  \hat{ \bf y }) $,
${\bf a}_{2}= (a/2) \, (\hat{ \bf x }  +  \hat{ \bf y }) $,
${\bf a}_{3}= (a/2) \, (\hat{\bf x}  +  \hat{ \bf z }) $,
and  primitive reciprocal vectors
${\bf b}_{1}= (2 \pi/a) ( \hat{\bf x} - \hat{\bf y} - \hat{\bf z}  )$,
${\bf b}_{2}= (2 \pi/a) ( \hat{\bf x} + \hat{\bf y} - \hat{\bf z}  )$,
${\bf b}_{3}= (4 \pi/a) \hat{\bf z}  $,
with $a \approx 3.61$ $\AA$ and $k_{F} \approx 1.36$ $\AA^{-1}$.
Then, for 
${\bf q}= {\bf k} - {\bf k}^{\prime}$, with
$ {\bf k}$ and  ${\bf k}^{\prime}$ on the Fermi surface, 
it follows that  
\begin{equation}
|{\bf q}| \leq 2 k_{F} < |{\bf b}_{1} |, |{\bf b}_{2} |
\;  ,
\end{equation}
implying that the only acceptable choice 
is $n_{1}=n_{2}=0$, namely, ${\bf g}_{n_{1}n_{2}}={\bf 0}$.
Under these conditions,
${\bf q}_{\rm in}={\bf 0}$ and also ${\bf q}_{\parallel}={\bf 0}$,
because 
${\bf q} \cdot {\bf a}_{1} =0 $
and  ${\bf q} \cdot {\bf a}_{2} =0 $ simultaneously; thus,
 ${\bf q}= {\bf q}_{\perp}$.
 Then,
 the argument of the interference factor 
${\cal F}_{N_{3}}(\xi_{3})   $
is
\begin{equation}
\xi_{3} = {\bf q}_{\perp} \cdot {\bf a}_{3} =
q_{z} d
\;  ,
\label{eq:xi_3}
\end{equation}
with $d$ being the distance between consecutive monolayers.
As a consequence,
Eq.~(\ref{eq:potentialsquared}) becomes
\begin{equation}
{\cal J} ({\bf q})=
{\cal J}^{(\perp)}  ( q_{z} )
\, \delta_{ {\bf q}_{\parallel},{\bf 0} }
\;  ,
\label{eq:potentialsquared_symmetry}
\end{equation}
where
\begin{equation}
{\cal J}^{(\perp)}  ( q_{z} )
=
{\cal F}_{N_{3}}( q_{z} d )
\,
{\cal J}^{(0)}  ( q_{z} \hat{\bf z} )
\;  ,
\label{eq:potentialsquared_perp}
\end{equation}
with
\begin{equation}
{\cal J}^{(0)}   ( q_{z} \hat{\bf z} ) =
\left|
\int_{\cal C}  \frac{d^{3} r }{v}  V({\bf r}) 
e^{-i  q_{z} z } 
\right|^{2} 
\;  
\label{eq:potentialsquared_perp_0}
\end{equation}
[cf. Eq.~(\ref{eq:potentialsquared_0})].
Equations~(\ref{eq:potentialsquared_symmetry})--(\ref{eq:potentialsquared_perp_0}) 
 give just a Bragg scattering contribution and zero resistivity
as $N_{3} \rightarrow \infty$,
but fall short of that singular behavior for $N_{3}$ finite.
Based on the preceding analysis, the Bragg scattering contributions,
${\cal J}_{\rm Bragg}({\bf q})$
and ${\cal J}^{(\perp)}_{\rm Bragg}  ( q_{z} )$,
are defined to be the  
 asymptotic  forms of the functions ${\cal J}({\bf q})$
and $ {\cal J}^{(\perp)}  ( q_{z} )$ 
as $ N_{3} \rightarrow \infty $;
thus, they are related by Eq.~(\ref{eq:potentialsquared_symmetry}), 
with
\begin{equation} 
{\cal J}^{(\perp)}_{\rm Bragg}  ( q_{z} )
 = 
\frac{2\pi}{N_{3}}
 \sum_{n_{3}=-\infty}^{\infty}
\delta( q_{z} d  
- 2\pi n_{3})
\, 
{\cal J}^{(0)} \left( \frac{2 \pi n_{3}}{d} \hat{\bf z} \right)
\;  ;
\label{eq:Mlimit}
\end{equation}
in particular,
\begin{equation} 
\left[
{\cal J}_{\rm Bragg}  ( {\bf q}_{\parallel}= {\bf 0},q_{z} )
 \right]_{ |q_{z}| < 2 \pi/d
} 
=
\left[
{\cal J}^{(\perp)}_{\rm Bragg}  ( q_{z} )
 \right]_{ |q_{z}| < 2 \pi/d
} 
 =
 \frac{2\pi }{ N_{3} } \delta( q_{z} d )
\,    {\cal J}^{(0)}( {\bf 0} )
\; ,
\label{eq:J_Bragg}
\end{equation}
an expression that will be important in the derivation of the electrical
resistivity in the next section.

\section{Finite Electrical Resistivity}
\label{sec:resistivity}

Equation~(\ref{eq:conductivity_sum}) will give a nonzero electrical 
resistivity only when the t matrix develops
a nonzero imaginary part, i.e., 
$\Delta_{ {\bf k}_{\parallel}
k_{z}} (\epsilon) \neq 0$.
This imaginary part may arise in the process of replacing 
discrete sums by energy integrals
[according to the rule defined by Eq.~(\ref{eq:cont_longitudinal})],
if the self-energy acquires an analytic structure characterized by a branch
cut that is effectively generated by the merging of the
discrete poles of the discrete self-energy.
However, the scattering by a periodic lattice 
generates constructive interference in discrete directions 
(represented by $\delta$ functions); this Bragg
scattering fails to produce an imaginary self-energy.

Due to the directional nature of Bragg scattering,
the term 
${\cal J}^{(\perp)}_{\rm Bragg}( k_{z}- k_{z}^{\prime})$ given in
Eq.~(\ref{eq:Mlimit})
does not contribute to the sum of Eq.~(\ref{eq:Delta}), namely, 
\begin{displaymath}
- 
\sum_{{\bf k}_{\parallel}^{\prime},
k_{z}^{\prime}}
{\cal J}_{\rm Bragg}({\bf k}-{\bf k}^{\prime}) \,
{\rm Im} [ G^{0}_{{\bf
k^{\prime}}_{\parallel}
k_{z}^{\prime}}(\epsilon) ]
=0
\; .
\end{displaymath}

Then, we are led to
a resolution of the function ${\cal J}({\bf q})$ of Eq.~(\ref{eq:J})
 into two parts,
\begin{eqnarray}
{\cal J}({\bf q}) 
& = &
{\cal J}_{\rm Bragg}({\bf q})
+{\cal J}_{\rm diff}({\bf q})
\nonumber \\
& = &
\left[
{\cal J}^{(\perp)}_{\rm Bragg}( q_{z} )
+{\cal J}^{(\perp)}_{\rm diff}( q_{z} )
\right]
\,
\delta_{ {\bf q}_{\parallel},{\bf 0} }
\;  ,
\label{eq:J_resolution}
\end{eqnarray}
corresponding to Bragg scattering, given by Eq.~(\ref{eq:Mlimit}),
 and the remainder,
which we identify as background diffusive scattering.
This procedure, which is based on the analysis of
Sec.~\ref{sec:periodicity},
amounts to isolating the 
diffusive part of the scattering
$
{\cal J}_{\rm diff}({\bf q}) = {\cal J}({\bf q}) - {\cal
J}_{\rm Bragg}({\bf q})
$, which  leads to a finite resistivity via the term
\begin{eqnarray}
\Delta({\bf k}_{\parallel}, k_{z};\epsilon) 
& = &
{\pi}
\sum_{{\bf k}_{\parallel}^{\prime},
k_{z}^{\prime}}
{\cal J}_{\rm diff}({\bf k}-{\bf k}^{\prime})
\delta(\epsilon
-\epsilon_{{\bf k}_{\parallel}^{\prime} }-\epsilon_{ k_{z}^{\prime}})
 \nonumber
\\
 & =  & 
{\pi}
\sum_{
k_{z}^{\prime} }
{\cal J}^{(\perp)}_{\rm diff}(k_{z} -k_{z}^{\prime})
\delta(\epsilon
-\epsilon_{{\bf k}_{\parallel} }-\epsilon_{ k_{z}^{\prime}})
\;  .
\label{eq:Delta2}
\end{eqnarray}
 Equation~(\ref{eq:Delta2})
 can be further simplified by either applying 
the longitudinal
continuum approximation, Eq.~(\ref{eq:cont_longitudinal}) 
with respect to $z^{\prime}$, or
explicitly rewriting the $\delta$ function as
\begin{eqnarray}
\delta(\epsilon -\epsilon_{ {\bf k} }    )
& = &
2 \pi g^{(\perp)} (\epsilon -\epsilon_{ {\bf k}_{\parallel} }    )
\,
\frac{1}{2} 
\left[
\delta
\mbox{\boldmath\large  $\left(  \right.$ } \! \!
k_{z} - k( \epsilon -\epsilon_{ {\bf k}_{\parallel} }    )
 \!  \mbox{\boldmath\large  $ \left.  \right)$ }
+
\delta
\mbox{\boldmath\large  $\left(  \right.$ } \! \!
k_{z} + k( \epsilon - \epsilon_{ {\bf k}_{\parallel} }    )
 \!  \mbox{\boldmath\large  $ \left.  \right)$ }
\right]
\nonumber \\
& = &
\frac{L}{2}  g^{(\perp)} (\epsilon -\epsilon_{ {\bf k}_{\parallel} }    )
\left[
\delta_{ k_{z},  k( \epsilon -\epsilon_{ {\bf k}_{\parallel} }) }    
+
\delta_{ k_{z}, - k( \epsilon -\epsilon_{ {\bf k}_{\parallel} }) }    
\right]
\;  .
\label{eq:delta_as_density}
\end{eqnarray}

Then,
\begin{equation}
\Delta({\bf k}_{\parallel}, k_{z};\epsilon) 
  =  
\frac{\pi L}{2}
 g_{\perp}
(\epsilon - \epsilon_{{\bf k}_{\parallel} })
\left[
{\cal J}_{\rm diff}^{(\perp)}
\mbox{\boldmath\large  $\left(  \right.$ } \! \!
k_{z} - k(\epsilon - \epsilon_{{\bf k}_{\parallel} })
\mbox{\boldmath\large  $\left.  \right)$ }  \!
+
{\cal J}_{\rm diff}^{(\perp)}
\mbox{\boldmath\large  $\left(  \right.$ } \! \!
k_{z} + k(\epsilon - \epsilon_{{\bf k}_{\parallel} })
\mbox{\boldmath\large  $\left.  \right)$ }  \!
\right]
\;  .
\label{eq:Delta3}
\end{equation}
Finally, for the
calculation of the conductivity, 
Eqs.~(\ref{eq:conductivity_sum}) and (\ref{eq:conductivity_integral}) 
dictate that Eq.~(\ref{eq:Delta3})
be evaluated for electrons on the Fermi surface, for which
the conditions $|k_{z}| = k(\epsilon_{F} - 
\epsilon_{{\bf k}_{\parallel} })$
and ${ k}_{\parallel}  = k(\epsilon_{F} - \epsilon_{ k_{z} })$
apply; then,
\begin{equation}
\Delta
\mbox{\boldmath\large  $\left(  \right.$ } \! \!
k(\epsilon_{F} - \epsilon_{ k_{z} }), k_{z}; \epsilon_{F} 
\mbox{\boldmath\large  $\left.  \right)$ }  \!
 = 
\frac{\pi L}{2} 
g_{\perp}(\epsilon_{k_{z}} )
\,
\left[
{\cal J}^{(\perp)}_{\rm diff}(0)
+ {\cal J}^{(\perp)}_{\rm diff}( 2 k_{z}  ) 
\right]
\; .
\label{eq:DeltaFermi}
\end{equation}
Substitution of  Eq.~(\ref{eq:DeltaFermi})
into Eq.~(\ref{eq:conductivity_integral}) leads to
 \begin{equation}
\sigma
 = 
\frac{e^{2} }{h}
\frac{2}{ \pi L}
\int^{\epsilon_{F}}_{0}
d  \epsilon_{ k_{z} } \,
 \frac{
\epsilon_{F }  - \epsilon_{ k_{z} } }{
{\cal J}^{(\perp)}_{\rm diff}( 0) 
+  {\cal J}^{(\perp)}_{\rm diff}( 2 k_{z}) } 
\; ,
\label{eq:conductivity_exact}
\end{equation}
where the properties 
${\cal J}^{(\perp)}_{\rm diff}( \pm 2 k_{z})  
=
\left[ {\cal J}^{(\perp)}_{\rm diff}( 2 k_{z})  \right]^{*}$
have been applied. Finally, the term
 $  {\cal J}^{(\perp)}_{\rm diff}( 2 k_{z})  $ 
can be approximated with its value 
 $  {\cal J}^{(\perp)}_{\rm diff}( 0)  =
   {\cal J}_{\rm diff}( {\bf 0})  $,
 because it is partially suppressed 
by the numerator as $k_{z}$ approaches $k_{F}$
(in fact, the exponential in the corresponding  Fourier integral
does not complete one entire cycle even as $z$ approaches $d$ and
$k_{z}$ approaches $k_{F}$ and is approximated as taking the value one);
then,
\begin{equation}
\sigma \approx 
\frac{e^{2}}{h}
\, \frac{ \epsilon_{F}^{2} }{ 2 \pi L{\cal J}_{\rm diff}( {\bf 0})}
\; .
\label{eq:conductivity_conclusion}
\end{equation}

In order to evaluate the final conductivity
expression of Eq.~(\ref{eq:conductivity_conclusion}), the value of 
\begin{equation}
{\cal J}_{\rm diff}({\bf 0})
=
\lim_{q_{z} \rightarrow 0}
\left[ {\cal J} \left(
{\bf q}_{\parallel}={\bf 0}, q_{z} \right)
 - {\cal J}_{\rm Bragg} 
\left( {\bf q}_{\parallel}= {\bf 0}, q_{z} \right)
\right]
\; 
\label{eq:Jdiff_limit}  
\end{equation}
is required.
This can be obtained from the interference factor,
 Eq.~(\ref{eq:FN}), which can be approximated for $N_{3}$ large and 
$\xi_{3}$ small via the expansion
\begin{eqnarray} 
{\cal F}_{N_{3}}(\xi_{3}) 
& = &
\frac{1}{N_{3}^{2}}
\frac{  \sin^{2} ( N_{3} \xi_{3}/ 2) }{ (\xi_{3}/2)^{2}  } 
\left[ 1 +  \frac{2}{3!}
\left( \frac{\xi_{3}}{2} \right)^{2}  \right] + O(\xi_{3}^{2})
\nonumber
\\
& \sim &
\frac{2\pi}{N_{3}}\delta(\xi_{3}) + \frac{1}{ 3 N_{3}^{2} }
\sin^{2} \left( \frac{N_{3}\xi_{3}}{2}
\right) + O(\xi_{3}^{2})
\;  ,
\label{eq:trivialexp}
\end{eqnarray}
where the denominator has been expanded in power series of
$\xi_{3}/2$ and Eq.~(\ref{eq:Mlimit_aux}) has been applied to
provide the asymptotic form of ${\cal F}_{N_{3}}(\xi_{3})$.
The derivation above focuses directly on the $\xi_{3} \rightarrow 0$ limit
and does not emphasize the resolution of the spectrum into
infinitely many Bragg peaks, each of which has a long 
tail that  contributes to the background diffusive scattering.
Instead,
a more illuminating approach would be to display all the Bragg peaks 
explicitly from the start
by considering the limit $\xi_{3}  \rightarrow 0$
of the simple fraction expansion of Eq.~(\ref{eq:exp_simple_fractions}), 
namely,
\begin{equation}
\lim_{\xi_{3} \rightarrow 0}
\left[
\frac{1}{\sin^2 (\xi_{3}/2)^{2}} -
\frac{1}{ (\xi_{3}/2)^{2} }
\right]
=
\;
 \frac{2}{\pi^2} \, 
\sum_{n=1}^{\infty}
\frac{1}{n^2}
= \frac{2}{\pi^2} \, \zeta (2) = \frac{1}{3} 
\;  ,
\label{eq:Riemann} 
\end{equation}
a result that is 
in agreement with Eqs.~(\ref{eq:Jdiff_limit}) and
(\ref{eq:trivialexp}).
In the formulas above,
in the limit $M\rightarrow\infty$, the factor
$\sin^{2}(M \xi_{3}/2 )$ oscillates fast about its average value
$1/2$, which would therefore be effectively achieved in
all physical measurements; thus,
one concludes that Eq.~(\ref{eq:trivialexp}) is simplified to
\begin{equation} 
{\cal F}_{N_{3}}(\xi_{3})
 = \frac{2\pi}{N_{3}}\delta(\xi_{3})+\frac{1}{ 6 N_{3}^{2} }
+ O(\xi_{3}^{2})
\;  .
\label{eq:FM_limit}
\end{equation}
Notice the characteristic appearance of the Bernoulli number $B_{2}=1/6$,
associated with either the expansion of Eq.~(\ref{eq:trivialexp})
or with the value $\zeta(2)= B_{2} \pi^2$
of the Riemann $\zeta$ function in 
Eq.~(\ref{eq:Riemann}).
Then, for $\xi_{3}= q_{z} d$ small, 
Eqs.~(\ref{eq:potentialsquared_symmetry}),
(\ref{eq:potentialsquared_perp}),
(\ref{eq:J_Bragg}), and
 (\ref{eq:FM_limit}) imply that
\begin{eqnarray} 
{\cal J} ({\bf q}_{\parallel}={\bf 0}, q_{z} )  
& = & 
{\cal F}_{N_{3}}( q_{z} d ) \,
{\cal J}^{(0)}( {\bf 0} )
\nonumber \\
& = & 
\left[
{\cal J}_{\rm Bragg}({\bf q}_{\parallel}={\bf 0}, q_{z}  ) 
\right]_{ |q_{z}| < 2 \pi/d}
+  \frac{1}{6N_{3}^{2}} {\cal J}^{(0)}( {\bf 0} )
+ O(q_{z}^{2})
\; ;
\label{eq:JM_limit}
\end{eqnarray}
 therefore, from Eqs.~(\ref{eq:potentialsquared_perp_0})
and (\ref{eq:Jdiff_limit}), and from $N_{3} \approx M$, 
the diffusive part of ${\cal J}({\bf 0})$ for $M$ monolayers is
\begin{equation}
{\cal J}_{\rm diff}({\bf 0}) =
\frac{1}{6 M^{2}}
\,  {\cal J}^{(0)} ({\bf 0})
=\frac{1}{6 M^{2} }  \left\langle V  \right\rangle^{2}
\;  ,
\label{eq:Jdiff0}
\end{equation}
with 
\begin{equation}
\left\langle V \right\rangle =
\int_{\cal C}  \frac{d^{3} r }{v}  V({\bf r}) 
\; 
\label{eq:average_potential}
\end{equation}
being the average potential.

Finally, replacing Eqs.~(\ref{eq:Jdiff0}) and (\ref{eq:average_potential})
into Eq.~(\ref{eq:conductivity_conclusion}), 
and  recalling that $L \approx M d$, we get a remarkably simple 
expression for the resistivity,
\begin{equation}
\rho
\approx \frac{\pi}{3} \frac{h}{e^{2}} \, 
\frac{ \Gamma^{2}\,  d}{M}
\label{eq:resistivity_conclusion}
\;  ,
\end{equation}
where
\begin{equation}
\Gamma=
\frac{1}{\epsilon_{F} } 
\int_{\cal C}  \frac{d^{3} r }{v}  V({\bf r}) 
= \;
\left\langle  \frac{V({\bf r})}{\epsilon_{F}}  \right\rangle
\;  ,
\label{eq:averageV}
\end{equation}
which is the average potential  relative to the Fermi energy,
is a dimensionless  parameter 
characterizing the relative strength of the periodic potential.

Equation~(\ref{eq:resistivity_conclusion}) can be evaluated numerically
by introducing a natural atomic resistivity
\begin{equation} 
\rho_{0}=\frac{h}{e^{2}} (1 \, \AA)
\approx 258 
\; {\rm \mu\Omega \, cm}
\; ,
\label{eq:resistivity_zero} 
\end{equation}
whence
\begin{equation}
\rho
\approx \frac{270 \,  \Gamma^{2} \, d[\AA] }{M} 
\; \;
{\rm \mu\Omega \, cm}
\label{eq:resistivity_conclusion_numbers}
\;  .
\end{equation}
Equation~(\ref{eq:resistivity_conclusion_numbers})  
 summarizes one of the main results of this paper.
It displays  a characteristic inverse proportionality 
with respect to the number $M$ of monolayers. 
In addition, notice that:

(i) The period  $d$ is {\em not\/}
 a free parameter, as it is uniquely determined 
from the crystal structure.

(ii)
As the atomic length scale
$d[\AA]$
is always of the order of unity,
the largest variations in the resistivity scale of
Eq.~(\ref{eq:resistivity_conclusion_numbers})
will come from the dimensionless 
parameter $\Gamma$.

(ii)
 $\Gamma$ is indeed  the {\em only\/}
 free parameter  within the framework
of approximations used in this model.

(iii)
 The value of $\Gamma$
can be independently estimated from calculations of cohesive energy.

For example, for copper,
reasonable estimates are provided by
the following values of the relevant parameters~\cite{sei:40}:
$d \approx 1.8$ $\AA$
 and $\Gamma \approx 2/7$;
 then, the predicted CIP resistivity 
is approximately $40/M$ ${\rm \mu\Omega \, cm}$, a value
reasonably close to the 
 $116/M$ ${\rm \mu\Omega \, cm}$
 found by the {\em ab initio\/}
method~\cite{bla:99}.

\section{conclusions}
\label{sec:conclusions}

It is generally recognized that a perfect
 infinite periodic
structure leads to no electrical resistance
because of the condition of constructive interference or Bragg
 scattering.
In this paper, we have
shown that,
at sufficiently low temperatures,
for sufficiently clean
samples, and for outgoing boundary conditions,
finiteness of a metallic film 
will prevent the potential from being
 perfectly periodic, will produce an effective background diffusive
scattering, and will cause 
a size-dependent resistivity
 inversely proportional to the number of monolayers.
Our free-electron estimate is in close agreement with similar results from 
{\em ab initio\/} calculations.
It would be interesting to see if this resistivity can
be measured experimentally by properly
simulating outgoing (perfectly diffusive) boundary conditions.

As a coda, it is worth mentioning that the Boltzmann equation approach
with the conventional relaxation time approximation fails for transport
through a finite, yet otherwise perfect, film. The usual ansatz for the
nonequilibrium distribution function ~\cite{mah:90b}
$f({\bf k})=f_{0}(k)+ {\bf k} \cdot {\bf E} \, C(k) $
is not applicable, because the deviation from equilibrium depends on the
orientation of the electric field relative to the crystal axes of the film
as well as on the angle between ${\bf k}$ and ${\bf E}$.

\acknowledgments{
This work was supported in part (P.M.L.) 
by the Office of Naval Research
together with the Defense Advanced Research Projects Agency
(Grant No.
N00014-96-1-1207), the National Science Foundation (Grant No.
INT-9602192),
 and NATO
(Grant No. CRG 960340).}

\end{document}